\documentstyle[preprint,aps,epsf]{revtex}
\tighten
\begin{document}
\draft
\title{Vacuum Energy of $CP(1)$ Solitons}
\author{I.~G.~Moss$^{1}$\thanks{email: {\tt ian.moss@ncl.ac.uk}},N.~Shiiki
             and T.~Torii$^{2} $\thanks{email: 
            {\tt torii@resceu.s.u-tokyo.ac.jp }}}
\address{$1$.Department of Physics, University of Newcastle Upon Tyne, 
NE1 7RU U.K.}
\address{$2$. Research Center for the Early Universe, Graduate School of
Science, University of Tokyo, 3-17-1, Hongo,Bunkyo-ku, Tokyo, Japan }
\date{March 21 2001} \maketitle
\begin{abstract}
The vacuum energy of two $CP(1)$ solitons on a torus is computed numerically. 
A numerical technique for the zeta-function regularisation is proposed 
to remove the divergence of the vacuum energy. After performing the 
numerical regularisation, we observe the effect of the vacuum energy on the 
two-soliton configuration. 
\end{abstract}
\narrowtext

\section{Introduction}
Topological solitons arise as solutions of classical field equations and
corresdond to points where the classical energy attains degenerate local
minima. The classical energy of a topological soliton typically
depends on an integer which can be interpreted as the number of solitons,
and in some cases it is independent of the parameters which correspond to the
positions and the sizes of individual solitons. Therefore the quantum
corrections can play a dominant role in the interactions between solitons. 

Many techniques have been used for the numerical computation of the vacuum
energies of single solitons \cite{dashen,goldstone,cahill,christ}. These
methods rely heavily on the spherical symmetry which is usually implicit in
the single soliton solutions. In this paper, we propose a new technique which
does not rely on any symmetry and can therefore be applied to multiple soliton
solutions. 

We will evaluate the vacuum energy for the $CP(1)$ model in $d = 1+2$
dimensions. Previous work on single solitons in the $CP(1)$ model has used an
approach based on heat kernel coefficients and phase shifts
\cite{ian-soliton}. It was shown that the solitons are unstable to collapse
due to the quantum corrections. The $CP(1)$ sigma model also exhibits
multi-soliton solutions. We focus on the charge-$2$ case and study their
interaction due to quantum corrections.

The quantisation of $CP(1)$ solitons can be performed by following the
standard  perturbation techniques invented by Schwinger \cite{schwinger} and
developed in the 1970s by several authors (for a review, see
\cite{jackiw-quantum}). The regularisation will be performed by the
zeta-function technique, which we convert into a form that can be evaluated
numerically. We examine how the finite one-loop energy depends on parameters
such as the separation of the two solitons and their width.

\section{$CP(1)$ Sigma Model in $(2+1)$ Dimensions}

The $CP(1)$ sigma model consists of a single complex scalar field taking a 
value in the one dimensional complex projective plane $CP(1)$. 
The action of the $CP(1)$ sigma model is given by 
 \begin{equation}
       S = \int_{}^{} d^{3}x \; \frac{|\partial_{\mu} u|^{2}}{(1+|u|^{2})^{2}} 
       \label{2-action-1}
\end{equation}    
where $\mu$ is the spacetime index running over $0,1,2$ and $|u|^{2} = 
u\bar{u}$. The complex projective plane being mapped to a sphere 
stereographically, the $CP(1)$ model is equivalent to the $O(3)$ 
model. Let $u$ and $\{\phi^{a}\} , \;a=1,2,3$ be coordinates of the $CP(1)$ 
and $O(3)$ respectively. The stereographic coordinates are given by 
\begin{equation}
	(u_{1}, u_{2}) = \left(\frac{\phi^{1}}{1-\phi^{3}},
	\frac{\phi^{2}}{1-\phi^{3}}\right) 
\end{equation}  
with $u = u_{1} + iu_{2}$ and $(\phi^{a})^{2} =1 $. 
Expressing the action (\ref{2-action-1}) in terms of $\phi^{a}$, one can 
recover the $O(3)$ sigma model action 
\begin{equation}
	S = \frac{1}{4}\int_{}^{} d^{3}x \; (\partial_{\mu}\phi^{a})^{2} \; .
\end{equation} 

Topological soliton solutions for the $O(3)$, or equivalently $CP(1)$ model, 
were discovered by Belavin {\it et al.} \cite{belavin-metastable}. To be precise, 
their solutions are instantons in two-dimensional Euclidean spacetime, but at 
a classical level, they are the same as solitons in three-dimensional Minkowski 
spacetime. 

The static energy of the $O(3)$ model is given by 
\begin{equation}
	E = \frac{1}{4}\int_{}^{} d^{2}x \; (\partial_{i}\phi^{a})^{2}
\end{equation}  
where $i$ takes the values $1,2$. 
The finite energy condition requires a boundary condition 
\begin{equation}
	\partial_{i}\phi^{a} \rightarrow 0  \;\;\;\;\; {\rm as}  
	\;\; |{\bf x}| \rightarrow \infty \; .
	\label{2-boundary-1}
\end{equation}  
Without loss of generality, we can define the asymptotic value of $\phi^{a}$ 
as  
\begin{equation}
	 (\phi^{a})^{2} \to 1\;\;\;\;\; {\rm as }  \;\; 
	|{\bf x}| \rightarrow \infty  
	\label{2-boundary-2}
\end{equation} 
which satisfies (\ref{2-boundary-1}). This boundary condition compactifies
the space into a sphere. Since the field  space is also a sphere, the homotopy
group of the fields is $\pi_{2}(S^{2})  = Z$ and hence there are soliton
solutions.  The topological charge is given by 
\begin{equation}
	Q = \frac{1}{8\pi}\int_{}^{} d^{2}x \; \epsilon_{ij} \epsilon_{abc}
	\phi^{a}(\partial_{i}\phi^{b})(\partial_{j}\phi^{c}) \; .
\end{equation}
From the obvious identity 
\begin{equation}
	(\partial_{i}\phi^{a} - \epsilon_{ij}\epsilon_{abc}\phi^{b}
	\partial_{j}\phi^{c})^{2} \geq 0  \; , 
\end{equation} 
one can derive 
\begin{equation}
	(\partial_{i}\phi^{a})^{2} \geq \epsilon_{ij} \epsilon_{abc}
	\phi^{a}(\partial_{i}\phi^{b})(\partial_{j}\phi^{c}) \; .
\end{equation} 
This implies 
\begin{equation}
	E \geq 2\pi |Q|  \; .
\end{equation} 
The soliton solutions attain minimum energy in each topological sector 
and hence saturate the equality. Thus they satisfy the first order differential 
equation 
\begin{equation}
	\partial_{i}\phi^{a} - \epsilon_{ij}\epsilon_{abc}\phi^{b}
	\partial_{j}\phi^{c} = 0 \; . 
\end{equation} 
This equation turns out to be equivalent to the Cauchy-Riemann conditions in 
the $CP(1)$ version 
\begin{equation}
	\frac{\partial u_{1}}{\partial x} = \frac{\partial u_{2}}{\partial y},  
	\;\;\;\;\; 
	\frac{\partial u_{2}}{\partial x} = -\frac{\partial u_{1}}{\partial y}
	 \; . 
\end{equation}
The general solution is given by the analytic function 
\begin{equation}
	u_{0}(z) = \frac{(z-b_{1})\cdots (z-b_{m})}
	{(z-a_{1})\cdots (z-a_{n})}
	\label{2-solution}
\end{equation}
where $ z = x + iy$, and $a_{i}, (i=1,\cdots n)$ and $b_{j}, (j=1,\cdots m)$ 
are the complex parameters characterising the position and size of solitons 
respectively. Note that the only singularities the field $u$ can have are 
isolated poles, since they merely correspond to the north pole of a sphere in 
the target space of the $O(3)$ model. 

The degree of the mapping is equal to the topological charge of the solution 
(\ref{2-solution}), which is also equal to the number of solutions when 
expressing $z$ in terms of $u_{0}$, 
\begin{equation}
	Q = {\rm max}(m, n) \; .
\end{equation} 
$\bar{u}$ corresponds to the anti-soliton solution and gives an opposite charge 
of $u$.

\section{Charge-$2$ $CP(1)$ Solitons on the Torus}

For the sake of numerical work, we consider the $CP(1)$ model on the 
torus \cite{speight-lump}. This requires a slight modification of the
previous soliton solutions.  We shall assume that the field $u(z)$ on the
torus satisfies the periodic  boundary condition 
\begin{equation}
	u(z +1 + i) = u(z) 
\end{equation}
The fact that the only singularities of $u_{0}(z)$ are poles means 
$u_{0}(z)$ is a meromorphic function. Then  $u_{0}(z)$ can be represented 
as an elliptic function.  

In the charge-$2$ case, the solution is represented by the Weierstrass
function  (see appendix \ref{appendix:weier}). For
simplicity,  we shall assume that the two solitons are symmetric in their size
and location.  Then there are only two complex parameters required which are
the degrees  of freedom of the size and separation. Thus we may write 
\begin{equation}
	u_{0}(z) = \frac{1}{\alpha (\wp(z)+\rho )}
	\label{2-spesol}
\end{equation}
with complex parameters $\alpha$ and $\rho$. 

\begin{figure}
           \epsfxsize=20pc
           \hspace*{3cm}
            \epsffile{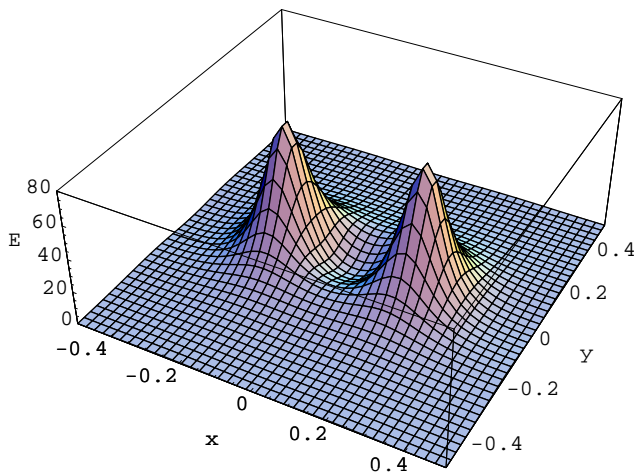}
	\caption{Energy density of charge $2$-soliton for 
	$\epsilon = 0.3$ and $w = 0.5$.}
	\protect\label{weier}
\end{figure}

Let the separation of the solitons (located at the poles) and the width be
$2\epsilon$ and $w$ respectively and restrict to real parameters. For  
(\ref{2-spesol}), one obtains  
\begin{equation}
	\rho   =  -\wp(i\epsilon),  \;\;\;\;\; \; 
	\alpha  =  \frac{2}{w \wp^{\prime }
	(i\epsilon)}  \; .
\end{equation}
In terms of $\epsilon $ and $w$, the solution 
(\ref{2-spesol}) becomes 
\begin{equation}
	u_{0}(z) = \frac{w \wp^{\prime  }(i\epsilon)}
	{2(\wp(z)-\wp(i\epsilon ))} \; .
	\label{cpsol}
\end{equation}
The energy density for this solution with $\epsilon = 0.3$ and $w = 0.5$ 
is plotted in fig.\ref{weier}. 

\section{Vacuum Energy of Charge-$2$ $CP(1)$ Solitons}

The static energy functional for the $CP(1)$ model on the torus 
is given by 
\begin{equation}
	E[u] = \int_{T_{2}}^{} d^{2}x \; \frac{|\partial_{i}u|^{2}}
	{(1+|u|^{2})^{2}}  \; .
\end{equation} 
The analytic expansion at $u_{0}$ is 
\begin{equation}
	E[u] = E_{0} + \int_{T_{2}}^{} d^{2}x \; G\,\bar{\xi}\Delta_{f}\xi + 
	\cdots 
\end{equation} 
where $\xi (z,\bar{z}) = u(z,\bar{z}) - u_{0}(z)$. The `metric' $G$ and the
fluctuation operator $\Delta_{f}$ are defined by  
\begin{equation}
	G = \frac{1}{(1+|u_{0}|^{2})^{2}}
\end{equation} 
and 
\begin{equation}
	\Delta_{f}\xi  \equiv \left[ -\nabla ^{2} + \frac{4\bar{u_{0}}
	(\partial _{\mu}u_{0})}{1+|u_{0}|^{2}}\partial ^{\mu} \right] \xi 
	\label{ch2-delf}
\end{equation} 
respectively. The computation of the vacuum energy depends on solving 
the eigenvalue problem 
\begin{equation}
	\Delta_f\chi_n=\lambda_n\chi_n\,.
	\label{2-omega}
\end{equation} 
Zero eigenvalues correspond to changes in the parameters, or moduli,
of the soliton solution. The quantisation of these parameters is treated
separately. Expanding $\xi $ in terms of the remaining eigenfunctions gives
an infinite sequence of (degeneracy two) oscillators with vacuum energy
$\lambda_n^{1/2}$. The total vacuum energy can be regularised using the
zeta-function scheme \cite{hawking-zeta,dowker-zeta}, 
\begin{equation} 	
E = E_{0} + \zeta \left(-\case1/2\right) 
\end{equation}
where the sum over non-zero eigenvalues
\begin{equation}
	\zeta(s) = \sum_{n=1}^{\infty} \lambda_{n}^{-s}  \; .
	\label{zeta}
\end{equation} 
gives the generalised Riemann zeta function for $s>1$. The value at $s=-1/2$
is uniquely determined by analytic continuation. 

\section{Numerical Regularisation}

We will now show how the zeta function regularisation scheme, which depends
on analytic continuation, can be converted into a numerical subtraction scheme.
Suppose that on average the eigenvalues approach the form  
\begin{equation}
	\widehat{\lambda }_{n} = an + b \; .
	\label{asymlam}
\end{equation}
Using (\ref{asymlam}), we can define the corresponding heat kernel 
\begin{equation}
	\widehat{K}(t) = \sum_{n=1}^{\infty} e^{-\hat{\lambda}_{n}t} \;. 
	\label{hatk}
\end{equation}
The behiour of the heat kernel as $t\to0$ gives information about the
eigenvalues at large values of $n$. We have
\begin{equation}
	\widehat{K}(t)= 
\frac{e^{-bt}}{1-e^{-at}} 
	  = \frac{1}{at} \sum_{n=0}^{\infty} 
b_{n}\left(\frac{b}{a}\right)
	  \frac{(at)^{n}}{n!}(-1)^{n}
\end{equation}
where $b_{n}(x)$ are the Bernoulli polynomials and the first three terms 
are given by 
\begin{equation}
	b_{0}(x) =1 \;\; ,\; \; b_{1}(x) 
	= x-\frac{1}{2} \; \; , \; \; 
	b_{2}(x) = x^{2}-x+\frac{1}{6} \; .
\end{equation}
For small $t$, the leading terms are  
\begin{equation}
	 \widehat{K}(t) =\frac{1}{at} - \left( \frac{b}{a}
	 -\frac{1}{2}\right) + O(t) \; .
	\label{leadkab}
\end{equation}
Comparing the order of $t$ in (\ref{4}) and (\ref{leadkab}), one can 
deduce 
\begin{equation}
	a  =  \frac{1}{B_{0}}  , \;\;\; \;\; \; b  =  \frac{1}{B_{0}}
	\left(-B_{1}+\frac{1}{2}\right)  \; .
	\label{ab}
\end{equation}
where $B_{0}$ and $B_{1}$ are the heat kernel coefficients for the operator
(\ref{cpsol}) calculated in appendix B, $B_0=1/(4\pi)$ and $B_1=2$. Thus,
$a=4\pi$ and $b=-6\pi$, with the result that 
\begin{equation}
	\widehat{\lambda }_{n} = 4\pi n-6\pi  \; .
	\label{lamhat}
\end{equation} 
We assign a zeta function to this eigenvalue
\begin{equation}
	\widehat{\zeta}(s) = \sum_{n=0}^{\infty} 
	\widehat{\lambda}_{n}^{-s} = 
	\left(\frac{1}{4\pi}\right)^{s}
	\zeta_H(s,\frac{1}{2})
	\label{zhatlam}
\end{equation}
where $\zeta_H(s,\frac{1}{2})$ is the generalised zeta function defined by
\begin{equation}
	\zeta_H(s,a) \equiv \sum_{n=0}^{\infty} (n+a)^{-s} \; .
	\label{gzet}
\end{equation}
Hurwitz has given a proof of the following formula 
\begin{eqnarray*}
	\zeta_H(s,a) & = & \frac{2\Gamma (1-s)}{(2\pi)^{1-s}}
	\left[\sin\left(\frac{s\pi }{2}\right)\sum_{n=1}^{\infty}n^{s-1}
	\cos(2n\pi a) \right. \\
	 & + & \left. \cos\left( \frac{s\pi }{2}\right)
	\sum_{n=1}^{\infty}n^{s-1}\sin (2n\pi a) \right] \; .
\end{eqnarray*}
Using this expression, one obtains 
\begin{equation}
	\widehat{\zeta } (s)  =  \left(\frac{1}{4\pi}
	\right)^{s}	\frac{2\Gamma(1-s)}{(2\pi)^{1-s}}
	 \sum_{n=1}^{\infty}n^{s-1}(-1)^{n}\sin\left(\frac{s\pi}{2}\right) \; .
\end{equation}
Setting $s=-1/2$ gives 
\begin{equation}
	\widehat{\zeta } \left(-\frac{1}{2}\right) = \frac{1}{2\sqrt{\pi }}
	\sum_{n=1}^{\infty}\frac{(-1)^{n+1}}{n^{3/2}} \; .
	\label{hatzs}
\end{equation} 
We can obtain a finite sum by subtracting the divergent terms,  
\begin{equation}
	\zeta_{{\rm reg}}(s) = \lim_{N\rightarrow \infty}\sum_{n=1}^{N}
	\left(\lambda_{n}^{-s}-\widehat{\lambda}_{n}^{-s}\right)   
         \label{numz}
\end{equation} 
where all the zero modes are removed from the eigenvalues. 
Then one can write the true zeta function as
\begin{equation}
	\zeta (s) = \zeta_{{\rm reg}}(s) +
\hat{\zeta}(s) \; . 
\end{equation}
Analytic continuation now implies 
\begin{equation}
	\zeta \left(-\frac{1}{2}\right) = \zeta_{{\rm reg}}\left(-\frac{1}{2}
	\right) + \hat{\zeta}\left(-\frac{1}{2}\right) \; .
	\label{numerical-zeta}
\end{equation}
The above expression allows us to evaluate the one-loop 
energy of the solitons numerically, assuming that (\ref{numz})
can be evaluated numerically. In practice, there are problems with the limit
$N\to\infty$ at $s=-1/2$ because the residuals $\lambda_n-\widehat\lambda_n$
do not vanish as $n\to\infty$. However, we shall see in later sections how
the difference does vanish `on average', allowing a result to be obtained. 

\section{Numerical Computation}

We use the Jacobi theta functions as a representation 
of the Weierstrass $ \wp$ function (see appendix \ref{appendix:weier})
\begin{equation}
	\wp (z) = \pi^{2} \left[\frac{\theta_{2}(0)
	\theta_{4}(0)\theta_{3}(\pi z)}
	{\theta_{1}(\pi z)}\right]^{2} \; .
	\label{numsol}
\end{equation}
Thus the charge $2$-soliton solution (\ref{cpsol}) can be 
written as 
\begin{equation}
	u_{0}(z) = i\pi w \frac{\theta_{3}^{2}(0)
	\theta_{2}(i\pi\epsilon)
	\theta_{4}(i\pi\epsilon)\theta_{1}^{2}(\pi z)}
	{\theta_{1}^{2}(i\pi \epsilon)\theta_{3}^{2}(\pi z)
	-\theta_{3}^{2}(i\pi \epsilon)\theta_{1}^{2}(\pi z)} \; .
	\label{u0num}
\end{equation}
In the following numerical computation, we shall use (\ref{u0num}) as 
a soliton solution.

We shall evaluate the approximate eigenvalues $\omega $ in 
(\ref{2-omega}) by the Rayleigh-Ritz variational method. 
This method introduces two principles. Firstly, the eigenfunctions 
are stationary configurations of the functional
\begin{equation}
	E[\xi ] = \frac{\int_{T_{2}}^{} d^{2}x \; G\bar{\xi}\Delta_{f}\xi }
	{\int_{T_{2}}^{} d^{2}x \; G|\xi |^{2}}\; .
	\label{2-eigenfunction}
\end{equation}  
Thus the eigenfunctions satisfy  
\begin{equation}
	\frac{\delta E[\xi  ]}{\delta \xi  } = 0 \; .
\end{equation} 
And secondly, trial functions which give stationary values of $E$ provide 
upper bounds for the eigenvalues (Hylleraas-Undheim theorem). 

The trial functions can be constructed as follows.  We take $N$ linearly
independent  functions $\phi_{n}$ parametrised by N variational parameters
$c_{n}$  and construct the trial function as their superposition
 \begin{equation}
	\xi = \sum_{n=1}^{N}c_{n}\phi_{n} \; .
\end{equation} 
Inserting into (\ref{2-eigenfunction}) one obtains the functional $E$ as a 
function of the $N$ variational parameters 
\begin{equation}
	E[c_{1}, c_{2}, \cdots c_{N}] = \frac{\sum_{m,n=1}^{N}
	\bar{c}_{m}c_{n}A_{mn}}
	{\sum_{m,n=1}^{N}\bar{c}_{m}c_{n}B_{mn}} 
	\label{functional-cn}
\end{equation} 
where we defined 
\begin{eqnarray}
	A_{mn} & = & \int_{T_{2}}^{} d^{2}x \; G\bar{\phi}_{m}\Delta_{f} 
	\phi_{n}
	\\
	B_{mn} & = & \int_{T_{2}}^{} d^{2}x \; G\bar{\phi}_{m}\phi_{n}  \; . 
\end{eqnarray}
Then, from the minimum principle, the upper bound of the $n$th eigenvalue 
is given by the stationary point of $E[c_{1}, c_{2}, \cdots c_{N}]$ 
with respect to $c_{n}$, i.e. $E$ satisfies  
\begin{equation}
	\frac{\partial E[c_{1}, c_{2}, \cdots c_{N}]}{\partial c_{n}}
	= 0 \; .
\end{equation}
Thus we obtain $N$ linear homogeneous equations for each upper 
bound of $E_{n}, \;\;\; (n=1,\cdots N)$ 
\begin{equation}
	\sum_{m=1}^{N}\bar{c}_{m}(A_{mn} - EB_{mn}) = 0 
\end{equation} 
and the problem is reduced to the $n$th-degree secular equation 
of the $N \times N$ matrix 
\begin{equation}
	{\rm det}({\bf A} - E{\bf B}) = 0 \; .
\end{equation}
Increasing in the number of basis states gives a lower upper bound of the
exact eigenvalue in each mode. 

We take the basis of trial functions $\phi_{\vec{k}}$ to be plane waves 
\begin{equation}
	\phi_{\vec{k}} = \exp\left\{\frac{i}{2}(\bar{k}z+k\bar{z})\right\}
	\label{2-plane}
\end{equation}
where $k=\frac{\pi}{L}(n+im)=2\pi(n+im)$ and $\vec{k}=(n,m) , \;\; 
n,m = -N, -N+1, \cdots , 0 , \cdots , N-1,N$. $n$ and $m$ are the mode 
numbers in the $x$ and $y$ directions respectively. With this basis, 
the functional  (\ref{functional-cn}) becomes 
\begin{equation}
	E[\bar{c}_{\vec{k}^{\prime}}, c_{\vec{k}}] = 
	\frac{\bar{c}_{\vec{k}^{\prime }}
	c_{\vec{k}}A_{\bar{k}^{\prime}k}}
	{B_{\bar{k}^{\prime}k}}
\end{equation}
where 
\begin{eqnarray}
	A_{\bar{k}^{\prime}k} & = & \int_{}^{}d\bar{z}dz 
	\frac{1}{(1+|u_{0}|^{2})^{2}}\bar{k}^{\prime }k
	\exp\left[\frac{i}{2}\{(\bar{k}^{\prime }
	-\bar{k})z+(k^{\prime }-k)\bar{z}\}\right] 
	 \\ 
	B_{\bar{k}^{\prime}k} & = & \int_{}^{} d\bar{z}dz 
	\frac{1}{(1+|u_{0}|^{2})^{2}}\exp\left[\frac{i}{2}
	\{(\bar{k}^{\prime }-\bar{k})z + (k^{\prime }-k)
	\bar{z}\}\right] \; .
\end{eqnarray} 
Therefore the upper bound of the spectrum can be computed 
from the secular equation of an $(2N+1) \times (2N+1)$ matrix 
\begin{equation}
	{\rm det}(A_{\bar{k}^{\prime}k} - EB_{\bar{k}^{\prime}k}) = 0 \; .
	\label{2-secular}
\end{equation} 
The spectrum will be more accurate as $N$ increases and in the limit 
$N \rightarrow \infty $, it is exact. 

To solve (\ref{2-secular}) numerically, we simply used the LAPACK 
(Linear Algebra Package) which provides routines for solving systems 
of simultaneous linear equations, least-square solutions of linear systems 
of equations, eigenvalue problems, and singular value problems. 
The file actually used is {\it chegv.f\/} for computing all eigenvalues of an 
Hermitian-definite generalised eigenproblem.

\section{Numerical Results}

\begin{figure}              
\epsfxsize=24pc            
\hspace*{2.5cm}        
\epsffile{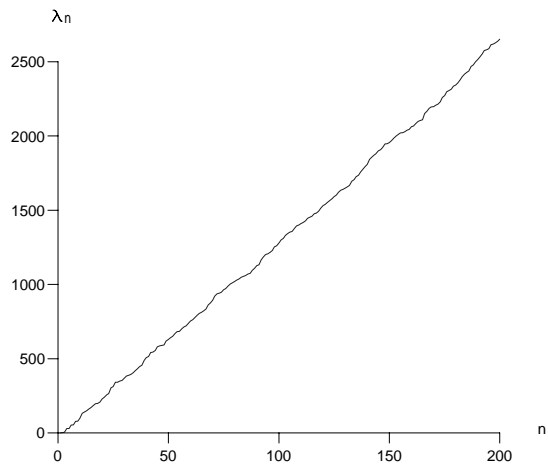} 	
\caption{nth-eigenvalue $\lambda_{n}$ as a
function of mode number $n$.} 	
\protect\label{eigenps}
\end{figure}

\begin{figure}             
\epsfxsize=26pc
           \hspace*{2.5cm}
            \epsffile{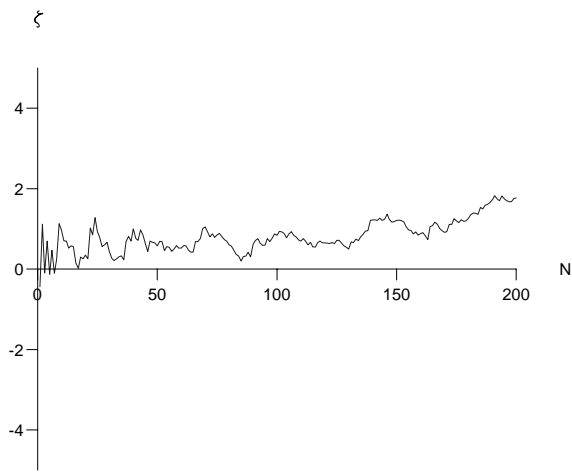}
	\caption{Regulated zeta function as a function of the 
	maximum mode number $N$.}
	\protect\label{zetaps}
\end{figure}

We have plotted the first 200 eigenvalues in fig.(\ref{eigenps}) as a function
of the mode number $n = (\bar{k}, k)$ for a typical background solution. The
agreement with the asymptotic formula (\ref{lamhat}) is quite striking.
Fig.(\ref{zetaps}) is a plot of the regularised zeta function as a function of
the number of modes $N$ up to which we took the sum of the eigenvalues. The
fluctuations in the data reflect the fact that the eigenvalues deviate from
the asymtotic formula. Although we can easily improve the accuracy of the
individual eigenvalues, these fluctuations have a significant influence in
preventing the convergence of the zeta-function in all of the cases which we
have examined. This problem has been seen before in numerical calculations of
zeta-functions \cite{balazs}. Averaging over $N$ has been proposed as a
solution to the problem and we use this in our calculation.

\begin{figure} 
          \epsfxsize=21pc
           \hspace*{3.0cm}
            \epsffile{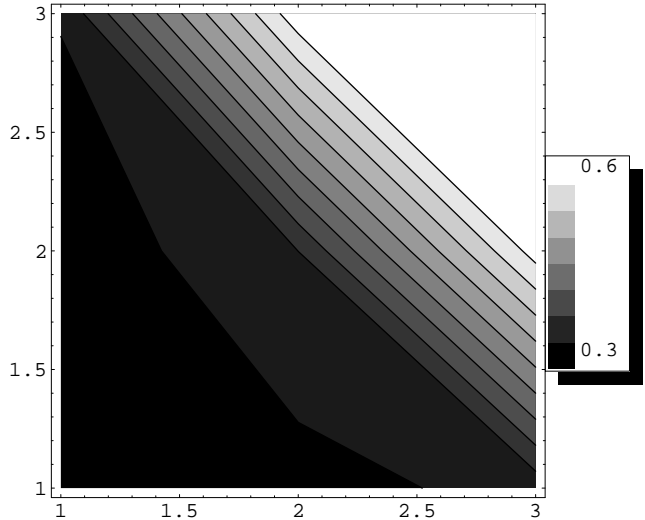}
	\caption{One loop energy of two separated solitons as a 
	function of $\epsilon$ and $w$.}
	\protect\label{contour}
\end{figure}

Fig.(\ref{contour}) is a contour plot of the average value 
of the zeta function on the parameter space of $\epsilon$ and 
$w$.  The data shows that the more spiky and closer two solitons are, 
the lower the vacuum energy is. Thus the classically stable two solitons
become unstable to collapse and merger when the one-loop quantum correction
is taken into consideration. 

\begin{figure}            
\epsfxsize=20pc            
\hspace*{2.5cm}
           \epsffile{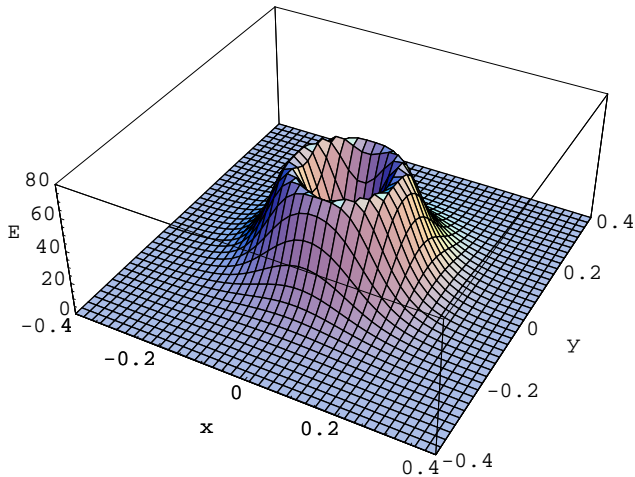}
	\caption{Energy density of a ring for $w=1.0$.}
	\protect\label{weierring}
\end{figure}

\begin{figure}
           \epsfxsize=20pc
           \hspace*{2.5cm}
            \epsffile{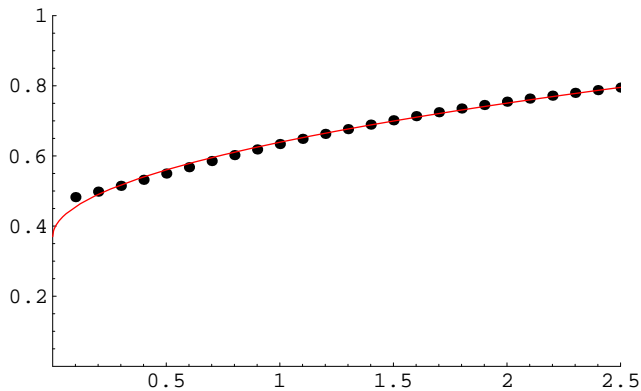}
	\caption{One loop energy of a ring as a function of 
	$w$.}
	\protect\label{fitring}
\end{figure}

As a special case of a charge-2 soliton, we also examined 
the one-loop energy for the solution 
\begin{equation}
	u_{0} (z) = \frac{w}{\wp(z)}  \; .
	\label{wring}
\end{equation} 
This solution exhibits a ring-shape energy configuration as is shown in 
fig.(\ref{weierring}). 
The numerical one-loop energy as a function of $w $ is given in 
fig.(\ref{fitring}). The plots denote the numerical values and the line 
is its interpolated function which is found to be 
\begin{equation}
	 0.269\sqrt{w} \; .
\end{equation} 
The rotational symmetry implies that this is also a case that can be analysed
using the phase shift technique \cite{ian-soliton}. Preliminary results
suggest an identical functional form for the vacuum energy, but the phase
shifts give a different overall scale \cite{noriko-th}.

\section{Conclusion }
In this paper, we have presented a numerical method for the computation of
finite one-loop energies for solitons and studied its effect on the
interaction of two $CP(1)$ solitons. The method does not require any symmetry
to be present in the problem, but it is expensive in computer time and
relies on the averaging out of residual terms in the eigenvalue
spectrum. Further analysis of these eigenvalue residuals, which should be
related to fundamental periods of the torus, would be useful and might improve
the technique.

The results are very satisfactory and show that initially separated solitons
are unstable to collapse and merger. Furthermore, the results are in
qualitative agreement with the fast and reliable phase shift method
\cite{ian-soliton} for a ring configuration, which has circular symmetry.

Interestingly, the vacuum energy is not the only effect on the dynamics of
two interacting solitons. There are additional forces arising from the
reduction of the quantum field theory to the $CP(1)$ moduli space
\cite{noriko}. In this example, an additional term is given by the Ricci
tensor of the moduli space, and gives a repulsive force between the two
solitons which is larger than the effect of the vacuum energy at very small
separations \cite{noriko-th}.

In principle, our methods are applicable to other solitons such as skyrmions, 
although moving from two to three dimensions would hughely increase the
computer time. The one-loop correction to the calculation of the skyrmion
mass is significant. For physical values of the pion decay constant, the
classical skyrmion has a mass about $50\%$ larger than a nucleon mass.
However, according to Moussallam and Kalafatis \cite{moussallam}, including
one loop energies within the framework of chiral perturbation theory, the
nucleon mass can be correctly predicted to within $20 \%$. Presumably, quantum
effects within chiral perturbation theory would also have an effect on the
force between two nucleons which could be calculated by the method presented
here. 

\appendix 

\section{Weierstrass Elliptic Functions}
\label{appendix:weier}

An elliptic function $f$ is a function such that it is doubly periodic with
two primitive periods  $2\omega_{1}, 2\omega_{2}$ whose ratio is not real,
i.e.  
\begin{equation}
	f(z + 2m\omega_{1} + 2n\omega_{2}) = f(z)
\end{equation} 
where $m, n$ are integers and 
\begin{equation}
	{\rm Im}\left(\frac{\omega_{1}}{\omega_{2}}\right) 
	\neq 0 \; . 
\end{equation}
Thus for an elliptic function $f(z)$, the $z$-plane can be 
partitioned into parallelograms with vertices $z_{0}+
2m\omega_{1}+2n\omega_{2}$. The only singularities of $f(z)$ allowed in a
period parallelogram are poles.

The weierstrass function $\wp $ is an even elliptic function of 
order $2$ with one double pole at $z = 2m\omega_{1} + 
2n\omega_{2}$ and defined by 
\begin{equation}
	\wp (z) = \wp (-z) = \frac{1}{z^{2}} + \sum_{m,n}^{}
	\left[\frac{1}{(z-m\omega_{1}-n\omega_{2})^{2}}
	- \frac{1}{(m\omega_{1}+n\omega_{2})^{2}}\right]
	\label{wp}
\end{equation} 
where $m, n$ takes all integers except for $m=n=0$. The series 
(\ref{wp}) converges everywhere except at the poles. 

The function $\wp (z)$ satisfies the differential equation 
\begin{equation}
	\left[\frac{d}{dz}\wp(z)\right]^{2} = 4\wp (z)^{3} 
	-g_{2}\wp (z) - g_{3} \equiv 4(\wp (z)-e_{1})
	(\wp (z)-e_{2})(\wp (z)-e_{3})
\end{equation} 
where $g_{2}$ and $g_{3}$ are invariants and determine the 
periods $\omega_{1}$ and $\omega_{2}$ as 
\begin{equation}
	g_{2} = 60 \sum_{m,n}^{}\frac{1}{(m\omega_{1}
	+n\omega_{2})^{4}} ,  \;\;\; 
	g_{3}=140 \sum_{m,n}^{}\frac{1}{(m\omega_{1}
	+n\omega_{2})^{6}} \; .
\end{equation} 
$e_{1}, e_{2}$ and $e_{3}$ are then given by 
\begin{equation}
	e_{1}=\wp (\omega_{1}), \;\; e_{2}=\wp (\omega_{2}), 
	\;\; e_{3}=\wp (\omega_{2})\; .
\end{equation}
with 
\begin{equation}
	e_{1}+e_{2}+e_{3}=0 ,  \;\; e_{1}e_{2}+e_{2}e_{3}
	+e_{3}e_{1}=-\frac{g_{2}}{4} ,  \;\; 
	e_{1}e_{2}e_{3}=\frac{g_{3}}{4} \; .
\end{equation} 

The Weierstrass function can be represented in terms of Jacobi 
theta functions as 
\begin{equation}
	\wp(z) = e_{j} + 
	\frac{\pi^{2}}{4\omega_{1}^{2}}\left[\frac{\theta^{\prime}
	_{1}(0)\theta_{j+1}(v)}
	{\theta_{j+1}(0)\theta_{1}(v)}\right]^{2} 
	\;\;\;\;\; j=1,2,3
	\label{weiertheta}
\end{equation}
where 
\begin{equation}
	v \equiv \frac{\pi z}{2\omega_{1}} 
\end{equation}
and the theta functions are defined by 
\begin{eqnarray*}
	\theta_{1}(z,q) & \equiv & \theta_{1}(z) = 
	2q^{\frac{1}{4}}\sum_{n=0}^{\infty}(-1)^{n}
	q^{n(n+1)}\sin\{(2n+1)z\}  \\
	\theta_{2}(z,q) & \equiv & \theta_{2}(z) = 
	2q^{\frac{1}{4}}\sum_{n=0}^{\infty}q^{n(n+1)}
	\cos\{(2n+1)z\}\\
            \theta_{3}(z,q) & \equiv & \theta_{3}(z) = 
            1+2\sum_{n=1}^{\infty}q^{n^{2}}\cos(2nz) \\
	 \theta_{4}(z,q) & \equiv & \theta_{4}(z) = 1+2
	 \sum_{n=1}^{\infty}(-1)^{n}
	 q^{n^{2}}\cos(2nz)  \; , 
\end{eqnarray*} 
with $q = e^{i\omega_{2}/\omega_{1}}$. 
$e_{1}, e_{2}$ and $e_{3}$ can also be expressed by the theta 
functions
\begin{equation}
	 \sqrt{e_{1}-e_{2}} = \frac{\pi}{2\omega_{1}}
	 \theta _{4}^{2}(0), \;\;  
	 \sqrt{e_{2}-e_{3}}  =  \frac{\pi}{2\omega_{1}}
	 \theta _{2}^{2}(0),  \;\;
	 \sqrt{e_{1}-e_{3}}  = \frac{\pi}{2\omega_{1}}
	 \theta _{3}^{2}(0) \; . 
	 \label{ej}
\end{equation} 

We adopt the theta-function representation for the weierstrass 
function since the theta functions converge rapidly for $n$ and the 
series are periodic.
For definiteness, we take the Lemniscatic case 
\begin{equation}
	\omega_{1}=-i\omega_{2}=\frac{1}{2} \;\;\;\;
	 {\rm and }  \;\;\;\;
	e_{2} = 0 \; .
\end{equation} 
Then from (\ref{ej}), we can determine numerical values 
of $e_{1}$ and $e_{3}$ as 
\begin{equation}
	e_{1}=-e_{3}=\pi^{2}\theta_{4}^{4}(0)=
	\frac{1}{8\pi}\Gamma\left(\frac{1}{4}\right)^{4}
	\sim  6.875 \; .
\end{equation} 
Choosing $j=2$ in (\ref{weiertheta}), one obtains 
\begin{equation}
	\wp (z) = \pi^{2} \left[\frac{\theta_{2}(0)
	\theta_{4}(0)\theta_{3}(\pi z)}
	{\theta_{1}(\pi z)}\right]^{2} \; .
\end{equation}

\section{Heat Kernel Coefficients}
\label{appendix:heat}

The small time expansion of a heat-kernel provides us a great deal 
of information about operators and their eigenvalues. This expansion 
was substantially developed in the $1960$s, and more of the 
mathematical details can be found in the original work and reviews 
for example \cite{gilkey-invariance,ian-quantum}. 

Consider the eigenvalue problem on a manifold ${\mit M}$  
\begin{equation}
	\Delta \phi = \lambda \phi
	\label{1}
\end{equation}
The field $\phi$ can have both spacetime, spinor and internal 
group indices, while $\Delta$ is the elliptic operator. The (integrated)
heat-kernel of the operator $\Delta$ is defined by   
\begin{equation}
 	K(t) = 
 	\sum_{n}^{}e^{-\lambda_{n}t}
 	\label{2}
\end{equation}
 We consider operators of the form 
\begin{equation}
 	-D^{2} + X
 	\label{3}
 \end{equation}
with a gauge covariant derivative $D=\nabla + A$ acting 
on fields which are associated with a representation of 
some given gauge group. For these operators, Gilkey has shown that the function 
$K(t)$ has an asymptotic expansion 
 \begin{equation}
 	K(t) \sim t^{-d/2}\sum_{n=0}^{\infty}B_{n}t^{n}
 	\label{4}
 \end{equation}
in $d$-dimensional space. In general, the $B_{n}$ coefficients depend on the 
operator, the geometry of the manifold and the boundary 
conditions satisfied by the fields $\phi$. 

The coefficients $B_{0}$ and $B_{1}$ for the $CP(1)$ 
model can be obtained by applying the general expressions. First, we rewrite
the fluctuation  operator obtained in \ref{ch2-delf} in a covariant form 
\begin{equation}
	-\partial^{\mu}\partial_{\mu} + 
	\frac{4\bar{u_{0}}(\partial_{\mu}u_{0})}
	{1+|u_{0}|^{2}}\partial^{\mu} =
	-2\left( D_{z}D_{\bar{z}}+D_{\bar{z}}D_{z}\right)
	- \frac{4|\partial_{z}u_{0}|^{2}}
	{(1+|u_{0}|^{2})^{2}}
	\label{cfluope}
\end{equation}
where $D_{z}$ is a covariant derivative defined by 
\begin{equation}
	D_{z} u = \partial_{z}u - 
	\frac{2\bar{u}_{0}(\partial_{z}u_{0})}
	{1+|u_{0}|^{2}}u \; .
	\label{covadel}
\end{equation}
Then the heat kernel coefficients for this operator are 
\begin{eqnarray}
	B_{0} & = & \frac{1}{4\pi} \int_{}^{} d^{2}x\,{\rm tr}(1)  
	\label{b0} \\
	B_{1} & = & \frac{1}{4\pi} \int_{}^{}d^{2}x \; (-X) = 
	\frac{1}{\pi} \int_{}^{} d^{2}x \; \frac{|\partial_{z}u_{0}|^{2}}
	{(1+|u_{0}|^{2})^{2}}  
	\label{b1}
\end{eqnarray}
Clearly, for a torus of unit area, $B_0=1/(4\pi)$. For a solution of the type 
\begin{equation}
	u_{0}(z) = \frac{\gamma}{z^{2}+\epsilon }\; ,
	\label{appen-cpsol}
\end{equation} 
one can show directly that 
\begin{equation}
	B_{1}  = 2 \; .
	\label{b0b1}
\end{equation} 
The same result can be seen to hold for the two-soliton solution on the
torus, because the integral (\ref{b1}) is the index of the mapping given by
$u$.



\begin{thebibliography}{99}
\bibitem{dashen}R. Dashen, B. Hasslache and A. Nevau, Phys. Rev D10 (1974) 4114
\bibitem{goldstone}J. Goldstone and R Jackiw, {\em Phys. Rev} {\bf D11} (1975)
1486 
\bibitem{cahill}K. Cahill {\em Phys. Lett.} {\bf B53} (1974) 174
\bibitem{christ}N. H. Christ and T. D. Lee  {\em Phys. Rev} {\bf D12} (1975)
1606
\bibitem{schwinger}J. Schwinger, {\em Phys. Rev.} {\bf 94} 1362 (1954)
\bibitem{jackiw-quantum} R. Jackiw, {\em Rev. Mod. Phys. } {\bf 49} 
(1977) 681
\bibitem{ian-soliton} I. G. Moss, {\em Phys. Lett. } {\bf B460} 
(1999) 103. 
\bibitem{belavin-metastable} A. A. Belavin and A. M. Polyakov, {\em JETP Lett. } 
{\bf 22}, No. 10 (1975) 245 
\bibitem{speight-lump} J. M. Speight, {\em Commun. Math. Phys. } 
{\bf 194} (1998) 513 
\bibitem{balazs} N. L. Balazs, C. Schmit and A. Voros, 
{\em J. Stat. Phys.} {\bf 46}: 5/6 (1987) 1067.
\bibitem{moussallam} B. Moussallam and D. Kalafatis, {\em Phys. Lett. } 
{\bf B272} (1991) 196 
\bibitem{gilkey-invariance} P. B. Gilkey, {\em Invariance Theory, the 
Heat Equation and the Atiyah-Singer Index Theorem} (Publish or Perish 
Inc., Wilmington Delaware, 1984)  
\bibitem{dewitt-dynamical} B. S. DeWitt, {\em Dynamical Theory of 
Groups and Fields} (Gordon and Breach, 1965) 
\bibitem{ian-quantum} I. G. Moss, {\em Quantum Theory, Black Holes 
and Inflation} (John Wiley and Sons Ltd, 1996) 
\bibitem{dowker-zeta} J. S. Dowker and R. Critchley, 
{\em Phys. Rev. } {\bf D16} (1977) 3390. 
\bibitem{hawking-zeta} S. W. Hawking, {\em Commun. Math. Phys.} 
{\bf 55} (1977) 133
\bibitem{noriko-th}N. Shiiki, PhD thesis ``Solitons and
black holes'' University of Newcastle upon Tyne (2000)
\bibitem{noriko}I. G. Moss and N. Shiiki, {\em Nucl. Phys.} {\bf B565} (2000)
345

\end{thebibliography}
\end{document}